\documentstyle[12pt]{article}
\begin{document}

\setlength{\textheight}{240mm}
\voffset=-25mm
\baselineskip=20pt plus 2pt
\begin{center}

{\large \bf The Energy for $2+1$ Dimensional Black Hole Solutions}\\
\vspace{5mm}
\vspace{5mm}
I-Ching Yang$^{\dag}$ \footnote{E-mail:icyang@nttu.edu.tw}
and Irina Radinschi$^{\ddag}$ \footnote{E-mail:iradinsc@phys.tuiasi.ro}

$^{\dag}$Department of Natural Science Education, \\
National Taitung University, \\
Taitung, Taiwan 950, Republic of China \\
and \\
$^{\ddag}$Department of Physics, "Gh. Asachi" Technical University, \\
Iasi, 6600, Romania 

\end{center}
\vspace{5mm}

\begin{center}
{\bf ABSTRACT}
\end{center}

The energy distributions of four 2+1 dimensional black hole solutions
were obtained by using the Einstein and M{\o}ller energy-momentum 
complexes.  While $r \rightarrow \infty$, the energy distributions 
of Virbhadra's solution for the Einstein-massless scalar equation 
becames $E_{\rm Ein} \sim \frac{\pi}{\kappa} (1-q) R^q$ and 
$E_{\rm M{\o}l} \sim -\frac{2\pi}{\kappa} q(1-q) R^q $, and the  
energy distributions of these three solutions become  
$E_{\rm Ein} \sim \frac{\pi \Lambda r^2}{\kappa}$ and
$E_{\rm M{\o}l} \sim -\frac{4\pi \Lambda r^2}{\kappa}$.

\vspace{2mm}
\noindent
{PACS No.:04.20.-q; 11.10.Kk}

\newpage

\section{INTRODUCTION}
One of the most interesting questions which remains unsolved in the general
theory of relativity is the energy-momentum localization. Numerous attempts
have been made in the past for a solution, and this question still attracts
considerable attention in the literature, and remains an important issue to
be settled. After the expression obtained by Einstein~\cite{1,2,3} for the
energy-momentum complexes, many physicist, such as Landau and Lifshitz~\cite{4},
Papapetrou~\cite{5}, Bergmann~\cite{6}, Weinberg~\cite{7} and M{\o}ller~\cite{2}
had given different definitions for the energy-momentum complexes. These 
definitions, except that of M{\o}ller, are restricted to evaluate the energy 
distribution in quasi-Cartesian coordinates. In his opinion Lessner~\cite{8} 
sustained that the M{\o}ller energy-momentum complex is an important 
concept of energy and momentum in general relativity. Some interesting 
results recently obtained~\cite{9,10,11,12} sustain the conclusion that the 
M{\o}ller energy-momentum complex gives reasonable results for many 
space-times.  Also, Cooperstock~\cite{13} gave his opinion that the energy 
and momentum are confined to the regions of non-vanishing 
energy-momentum tensor of the matter and all non-gravitational fields.

\section{BLACK HOLE SOLUTIONS IN $2+1$ DIMENSIONS}
In recent years the several gravity models in 2+1 dimensions have 
gained considerable attention~\cite{14}.  Due to an expectation is that 
the study of 2+1 dimensional theories would provide relevant information 
about the correponding theory in 3+1 dimensions.  In 2+1 dimensions the 
number of independent components of the Riemann curvature tensor and 
the Einstein tensor are the same, consequently the imposition of 
Einstein's equations in vacuum implies that the curvature rwnsor also 
vanishes.  Therefore, the space-time described by the vacuum solutions 
to Einstein equations in 2+1 dimensions are no gravitional wave and no 
interaction between masses, and the existence of black hole would be 
prevented~\cite{15}.  However, Ba{\~{n}}ados, Teitelboim, and Zanelli
(BTZ)~\cite{16} have discovered a black hole solution to the Einstein-Maxwell 
equations (with a negative cosmological constant) in 2+1 dimensions, which 
is characterized by mass, angular momentum, and charge parameters.  Because
of energy-momentum is a fundamental conserved quantity associated with a 
symmetry of space-time geometry. Thus, there are many study about that the  
energy of 2+1 dimensional black hole solutions in different gravity 
models~\cite{17}.
In this article, we will study the 
energy distribution of non-rotating black hole solutions in $2+1$ 
dimensions with Einstein and M{\o}ller energy-momentum complexes.  We use 
the geometrized units $(G=c=1)$ and adopt the signature of the metric 
with $(-,+,+)$. It follows the convention that the Latin indices run 
from $1$ to $2$ and the Greek indices run from $0$ to $2$ .  Now, we 
investigate these following 2+1 dimensional non-rotating black hole 
solutions with non-zero cosmological constant:  \\  
(i) Uncharged black hole solution~\cite{18}  \\
The solution is given by
\begin{equation}
ds^2=-(\Lambda r^2 -M)dt^2+(\Lambda r^2 -M)^{-1} dr^2 +r^2 d\theta^2   ,
\end{equation}
where $\Lambda$ is the cosmological constant.  \\
(ii) Charged black hole~\cite{18}  \\
This solution is expressed by the line element
\begin{equation}
ds^2=-(\Lambda r^2 -M -2Q^2 \ln \left( \frac{r}{r_+} \right))dt^2
+(\Lambda r^2 -M -2Q^2 \ln \left( \frac{r}{r_+} \right))^{-1} dr^2 
+r^2 d\theta^2   ,
\end{equation}
where $r_+ = \sqrt{M/ \Lambda}$.  \\
(iii) Coupling to a static scalar field~\cite{18} \\
The solution of 2+1 dimensional gravity field coupled to a static scalar
field is described by
\begin{equation}
ds^2= - \left( \frac{(r-2B)(B+r)^2 \Lambda}{r} \right) dt^2 +
\left( \frac{(r-2B)(B+r)^2 \Lambda}{r} \right)^{-1} dr^2 + r^2 d\theta^2
\end{equation}  \\
(iv) The static and circularly symmetric exact solution of 
the Einstein-massless scalar equation~\cite{19} \\
The static and circularly symmetric exact solution of the Einstein-massless 
scalar equation is given by
\begin{equation}
ds^2 = -B dt^2 + B^{-1} dr^2 + r^2 d\theta^2  ,
\end{equation}
where $B=(1-q)R^q$, $R=r/r_0$ and q stands for the scalar charge.

\section{ENERGY IN THE EINSTEIN PRESCRIPTION}
The well known energy-momentum complex of Einstein~\cite{1} is defined 
as
\begin{equation}
\Theta^{\nu}_{\mu} = \frac{1}{2\kappa} \frac{\partial}{\partial x^{\sigma}} 
H^{\nu \sigma}_{\mu} ,
\end{equation}
with superpotential
\begin{equation}
H^{\nu \sigma}_{\mu} = \frac{g_{\mu \rho}}{\sqrt{-g}}\frac{\partial}{\partial 
x^{\eta}} [(-g)(g^{\nu \rho}g^{\sigma \eta} - g^{\sigma \rho}g^{\nu \eta})] ,
\end{equation}
and the gravitational coupling constant $\kappa$.  Then we evaluate the 
energy and momentum by Einstein energy-momentum complex in quasi-Cartesian 
coordinates of 2+1 dimentional space-time, 
\begin{equation}
P_{\mu} = \frac{1}{2\kappa} \int \frac{\partial H^{oi}_{\mu}}{\partial x^i} d^2 x.
\end{equation}
and the energy component is obtained by using the Gauss theorem 
\begin{equation}
E_{\rm Ein} (r) = P_0 = \frac{1}{2\kappa} \int \frac{\partial H_0^{0i}}
{\partial x^i} dx dy   . 
\end{equation}
The metric of the non-rotating 
black hole solutions in 2+1 dimensions can be taken as in Cartesian coordinates
\begin{equation}
ds^2 = - v dt^2 + (\frac{x^2}{r^2}w+\frac{y^2}{r^2}) dx^2 
+ (\frac{2xy}{r^2}w-\frac{2xy}{r^2}) dxdy + (\frac{y^2}{r^2}w+\frac{x^2}{r^2}) dy^2   .
\end{equation}
Hence, the required nonvanishing components $H_0^{0i}$ of the Einstein 
energy-momentum complex are shown to be
\begin{eqnarray}
H_0^{01} & = & \frac{x}{r^2} \frac{v}{\sqrt{vw}} (1-w)  , \\
H_0^{02} & = & \frac{y}{r^2} \frac{v}{\sqrt{vw}} (1-w) .
\end{eqnarray}
Finally, we find the energy within a circle withe radius $r$ is
\begin{equation}
E_{\rm Ein} (r) = \frac{1}{2\kappa} \oint \frac{v}{\sqrt{vw}} (1-w) d\theta.
\end{equation}

\section{ENERGY IN THE M{\O}LLER PRESCRIPTION}
For another thing, the M{\o}ller energy-momentum complex~\cite{2} is given by
\begin{equation}
\Theta^{\nu}_{\mu} = \frac{1}{\kappa} \frac{\partial}{\partial x^{\rho}} \chi^{\nu \rho}_{\mu} ,
\end{equation}
with superpotential
\begin{equation}
\chi^{\nu \rho}_{\mu} = \sqrt{-g}(\frac{\partial g_{\mu \beta}}{\partial 
x^{\alpha}} - \frac{\partial g_{\mu \alpha}}{\partial x^{\beta}})
g^{\nu \alpha}g^{\rho \beta} .
\end{equation}
The energy and momentum using the M{\o}ller energy-momentum complex in 
spherical coordinates are calculated as
\begin{equation}
P_{\mu} = \frac{1}{\kappa} \int \frac{\partial \chi^{oi}_{\mu}}{\partial x^i} d^2 x.
\end{equation}
and the energy component is obtained by using the Gauss theorem 
\begin{equation}
E_{\rm M{\o}l} (r) = P_0 = \frac{1}{\kappa} \int \frac{\partial \chi_0^{0i}}
{\partial x^i} dr d \theta    . 
\end{equation}
For the 2+1 dimensional non-rotating black hole solutions in spherical 
coordinates, the metric can be expressed in the form
\begin{equation}
ds^2 = - v(r) dt^2 + w(r) dr^2 + r^2 d\theta^2   . 
\end{equation}
So, the only nonvanishing component of M{\o}ller energy-momentum complex is
\begin{equation}
\chi_0^{01} = - \frac{r}{\sqrt{vw}}\frac{\partial v}{\partial r}  .
\end{equation}
At last, the energy within a circle withe radius $r$ is obtained
\begin{equation}
E_{\rm M{\o}l} (r) = -\frac{1}{\kappa} \oint \frac{r}{\sqrt{vw}}
\frac{\partial v}{\partial r} d\theta  .
\end{equation}

\section{ENERGY DISTRIBUTION OF $2+1$ DIMENSIONAL BLACK HOLE SOLUTIONS}
To put these four space-time solutions(1)-(4) into equations (6) and 
(14), these cases furnish: \\
(i)Uncharged black hole solution: \\
\begin{equation}
E_{\rm Ein} =\frac{\pi (\Lambda r^2 -M-1)}{\kappa} , 
\end{equation}
and
\begin{equation}
E_{\rm M{\o}l} = - \frac{4\pi \Lambda r^2}{\kappa} .
\end{equation} \\
(ii)Charged black hole solution: \\
\begin{equation}
E_{\rm Ein} = \frac{\pi [\Lambda r^2 -M - 2Q^2 \ln ( \frac{r}{r_+} ) -1]}{\kappa} , 
\end{equation}
and
\begin{equation}
E_{\rm M{\o}l} = \frac{-4\pi (\Lambda r^2 - Q^2)}{\kappa} .
\end{equation} \\
(iii) Coupling to a static scalar field: \\
\begin{equation}
E_{\rm Ein} = \frac{\pi [\Lambda (r-2B)(B+r)^2 -r]}{\kappa r} , 
\end{equation}
and
\begin{equation}
E_{\rm M{\o}l} = \frac{-4\pi [\Lambda (B+r)(r^2 -Br+B^2)]}{\kappa r} .
\end{equation} \\
(iv)The static and circularly symmetric exact solution of the 
Einstein-massless scalar equation: \\
\begin{equation}
E_{\rm Ein} = \frac{\pi [(1-q)R^q-1]}{\kappa} , 
\end{equation}
and
\begin{equation}
E_{\rm M{\o}l} = \frac{-2\pi q(1-q)R^q}{\kappa} .
\end{equation}

\section{CONCLUSION}
The enery distributions for some $2+1$ dimensional black hole solutions
were computed with the Einstein and M{\o}ller energy-momentum complexes.  
In these four solutions the energy distributions of Einstein energy-momentum 
complex differ from M{\o}ller, and those each solution also differ with other 
three with Einstein's (or M{\o}ller's) prescription.  Specially, we found 
that all of the energy component of M{\o}ller energy-momentum complex are 
negative.  

As $r$ becomes larger, the energy distribution of Virbhadra's solution for the 
Einstein-massless scalar equation becomes
\begin{equation}
E_{\rm Ein} \sim \frac{\pi}{\kappa} (1-q) R^q
\end{equation}
and
\begin{equation}
E_{\rm M{\o}l} \sim -\frac{2\pi}{\kappa} q(1-q) R^q ,
\end{equation} 
and the energy distributons of these three solutions become 
\begin{equation}
E_{\rm Ein} \sim \frac{\pi \Lambda r^2}{\kappa}
\end{equation}
and
\begin{equation}
E_{\rm M{\o}l} \sim -\frac{4\pi \Lambda r^2}{\kappa}.
\end{equation} 
Because those solutions are not asymtotical flat, $E_{\rm Ein}$ and 
$E_{\rm m{\o}l}$ of all solutions divergence while $r \rightarrow \infty$. 
Nevertheless those energy-momentum complex still could be used to study 
the topology of black hole solutions. 

\begin{center}
{\bf Acknowledgements}
\end{center}
The authors would like to thank Prof. S. Deser, Prof. R. Jackiw, 
Prof. R. Mann and Dr. E.C. Vagenas for useful comments.  
I.-C. Yang also thanks the National Science Council of the Republic of 
China for financial support under the contract number NSC 92-2112-M-006-007

\end{document}